\documentclass[prd,superscriptaddress]{revtex4}
\usepackage{graphicx}
\usepackage[dvips]{hyperref}
\usepackage[T1]{fontenc}
\usepackage{amsmath,amssymb,mathbbol}
\usepackage{amsfonts}
\usepackage{euscript}
\usepackage{psfrag}
\usepackage{stmaryrd,latexsym}

\newcommand{\A}{\mathcal{A}}

\begin{document}
\title{Born-Infeld inspired bosonic action in a noncommutative geometry}
%\maketitle
\date{\today} 
\author{Emmanuel \surname{Seri\'e}}
\affiliation{Laboratoire de Physique Th\'eorique (UMR 8627),  Universit\'e Paris XI,  B\^atiment 210, 91405 Orsay Cedex, France}
\affiliation{Laboratoire de Physique Th\'eorique des Liquides (UMR 7600), Universit\'e Pierre-et-Marie-Curie, Tour 24, 4-\`eme \'etage, Bo\^{i}te 121,   4, Place Jussieu, 75005 Paris, France}
\author{Thierry \surname{Masson}}
\affiliation{Laboratoire de Physique Th\'eorique (UMR 8627),  Universit\'e Paris XI,  B\^atiment 210, 91405 Orsay Cedex, France}
\author{Richard \surname{Kerner}}
\affiliation{Laboratoire de Physique Th\'eorique des Liquides (UMR 7600), Universit\'e Pierre-et-Marie-Curie, Tour 24, 4-\`eme \'etage, Bo\^{i}te 121,   4, Place Jussieu, 75005 Paris, France}

\begin{abstract}
The Born-Infeld lagrangian for non-abelian gauge theory is adapted to the case of the  generalized gauge fields arising in non-commutative matrix geometry. Basic properties of static and time dependent solutions of the scalar sector of this model are investigated.
\end{abstract}
\pacs{11.10.St, 11.15.-q, 11.27.+d, 12.38.Lg, 14.80.Hv}
\maketitle
LPT-Orsay 04-65
%%%%%%%%%%%%%%%%%%%%%%%%%%%%%%%%%%%%%%%%%%%%%%%%%%%%%%%%%%%%%%%%%%%%%
\section*{Introduction}

In this article we propose to extend the non-abelian generalization of Born-Infeld\cite{born_infeld:34} lagrangian proposed in \cite{serie:03} to the noncommutative  geometry of matrix valued functions on a manifold  \cite{dubois-violette:90,dubois-violette:90:II}. It was shown that such realization of gauge principle contains not only the usual $SU(n)$ gauge field, but also a generalization of Higgs multiplet of scalar fields. The matrix realization of noncommutative geometry provides also a framework in which the calculus of a determinant can be naturally generalized.

It should be stressed that the action proposed in this paper provides a natural generalization of the so-called Dirac-Born-Infeld (DBI) action\cite{dirac:62,born_infeld:34}(see \cite{gibbons:98} and references therein) for non-abelian gauge theories. Such a generalization was also proposed by Tseytlin \cite{tseytlin:97} via dimensional reduction of pure non-abelian Born-Infeld lagrangian action computed with symmetric trace prescription. Our approach here is similar  but we propose to use noncommutative geometry of matrices instead of dimensional reduction and to use a  determinant in the tensor product\cite{serie:03} instead of the symmetric trace prescription. Our lagrangian is closed to the one proposed by Park~\cite{park:99}, but, as shown in our previous paper~\cite{serie:03}, the main difference being the use of hermitian generators of the Lie algebra and introduction of an additional quasi-complex structure.

Other  non-abelian Born-Infeld lagrangians with scalar fields were proposed in the context of study of classical solutions \cite{brihaye:01,dyadichev:02,brihaye:03} by simple addition of the usual lagrangian for Higgs fields to a standard  Born-Infeld lagrangian. 

Our interest is focused on the pure noncommutative scalar sector. We show that soliton-like solutions with finite energy can not be obtained with pure Higgs fields obeying this version of generalized Born-Infeld dynamics, in the case when the Higgs multiplet reduces to a single scalar $\varphi$. Such an anzatz is natural if spherical symmetry is imposed (further studies are in progress in order to treat correctly the spherical symmetry in the framework of matrix noncommutative geometry).

Next, we consider a time-dependent scalar field and its dynamics. The equations of motion are highly non-linear, and there is little hope to get genuine analytic solutions. Nevertheless we can say quite a lot about the qualitative behavior of solutions by exploring the phase space and spotting all singular points and curves. As it could be expected in a Born-Infeld-like theory, the non-divergent trajectories in the phase space $\varphi, \, \ \ u = \dot{\varphi}$ are confined within  certain limits, as a natural consequence of the existence of maximal field strength. This feature of the theory makes it particularly interesting for cosmological models with non-standard scalar fields as driving force for accelerated inflation.

\section{Gauge fields in noncommutative geometry}

We shall generalize here the "noncommutative Maxwell theory" developed in \cite{dubois-violette:90:II} which will be used as a framework for a Born-Infeld inspired lagrangian. Let's first recall the basis assumptions of the noncommutative geometry in a particular matrix realization which will be used here. We consider the algebra $\A = C^{\infty}(V) \otimes M_n(\mathbb{C})$ with the vector fields spanned by the derivations of $ C^{\infty}(V)$ and inner derivations of $ M_n(\mathbb{C})$. The differential algebra is generated by the basis of linear 1-forms acting on the derivations. We can consider $\A$ as a bimodule over itself. Then one fixes the gauge choosing a unitary element $e$ of $\A$, satisfying $h(e,e)=1$, with $h$ a hermitian structure on $\A$. Then any element of $\A$ can be written in the form $ \, em \, $ with $m \in \A$, and a connection on $\A$ can be defined as a map: 
\begin{align*}
  \nabla:  \A & \to  \Omega^{1}(\A)  &   
   em & \mapsto (\nabla e)\ m  + e \  dm
 \end{align*}
In the gauge $e$, a  connection can be completely characterized by an element
$\omega$ of $\Omega^{1}(\A)$:
\begin{align*}
\nabla e &= e \  \omega \ .
\end{align*}
One can also  decompose $\omega$ in vertical  and horizontal parts
\begin{align*}
  \omega &= \omega_h + \omega_v  & & \text{with}&
  \omega_h &= A  &
  \omega_v &= \theta + \phi \ .
\end{align*}
Here $A$ is an analog of the Yang-Mills connection, whereas $ \theta $  is the canonical  1-form of the matrix algebra, and plays the role   of preferred origin in the affine space of vertical connections.   It satisfies the equation:
  \begin{align*}
    d\theta +\theta \wedge \theta &=0
  \end{align*}
Then  $\phi$ is a tensorial form and can be  identified with scalar field multiplet.

 One can choose a local basis of derivations of $\A$: $\{e_{\mu},e_{a}\}$,  where for convenience $e_{\mu}$ are outer derivations of $C^{\infty}(V)$, and $e_{a}= ad(\lambda_{a})$, with $\{\lambda_{a}\}$ a basis of anti-hermitian  matrices of $M_n(\mathbb{C})$, are inner derivations.\\
 The dual basis will be denoted by  $\{ \theta^{\mu}, \theta^{a} \}$.  In this particular basis, we have:
\begin{align*}
   A &= A_{\mu} \theta^{\mu}  &
   \theta &= - \lambda_a \theta^a  &
   \phi &= \phi_a \theta^a 
 \end{align*}
If we choose the connection to be anti-hermitian, we can write    $ \phi = \phi_a^b \lambda_b \theta^a$. The curvature tensor associated with $\omega$ is :
\begin{align*}
  \Omega = d\omega + \omega \wedge \omega
\end{align*}
we can also define the field strength: 
\begin{align*}
  F&= dA + A \wedge A \ .
\end{align*}
Then one can identify:
\begin{align*}
  \Omega_{\mu \nu} &= F_{\mu \nu} & \Omega_{\mu a} &= D_{\mu} \phi_a\\
  \Omega_{a \mu } &= - D_{\mu} \phi_a &
  \Omega_{a b} &= [\phi_a,\phi_b] - C_{a b }^{c} \phi_c
\end{align*}
 $ C_{a b }^{c} $ are the structure constants in the $\{\lambda_a\}$ basis.

A gauge transformation is performed by the choice of a unitary element $U$ of $M_n(\mathbb{C})$, satisfying $ h(e U, e U ) = 1$. Then in the gauge $e'=e U$
\begin{align*}
  \omega' &= U^{-1} \omega U + U^{-1} dU
\end{align*}
$\theta$ is invariant under gauge transformations, then  $A$ and $\phi$  transform as follows:
\begin{align*}
  A' &=U^{-1}  A U + U^{-1} dU  &
  \phi' &=U^{-1}  \phi U 
\end{align*}
Taking into account that all forms appearing here are matrix-valued, it is quite natural to use the invariants provided by generalized  determinants in the construction of lagrangian densities. For more details, see J. Madore's book \cite{madore:99}.

\section{Noncommutative Born-Infeld Lagrangian}

The generalization proposed in our previous article\cite{serie:03} can be adapted in the noncommutative gauge theory framework. The lagrangian which we shall  consider is:
\begin{align}
 {\cal L} &= \sqrt{\det|g|} - \{|\det ( \mathbb{1} \otimes g + J \otimes \Omega |\}^{1/4n}
  \label{binc}
\end{align}
and $\Omega = \Omega_{\alpha \beta}\hat{L}^{\alpha \beta} $ with $\hat{L}^{\alpha \beta} $ the generators of the fundamental representation of $SO(4+n^2-1)$. $ \Omega_{\alpha \beta}$ are the components of curvature 2-form defined in previous section, and then are anti-hermitian elements of $M_n(\mathbb{C})$. $J$ is an element of $SL(2,\mathbb{C})$ of square $-\mathbb{1}$. 

Similar approach was suggested as early as in 1981 by Hagiwara \cite{hagiwara:81}, and developed later on by Park \cite{park:99}. Our improvement consisted in ensuring the hermiticity by incorporating quasi-complex structure generated by an extra matrix $J$,  and in taking into full account the discrete geometry of matrices.

The above lagrangian contains contributions coming from two types  of fields: the classical Yang-Mills potential, $ A = A_{\mu} \theta^{\mu}$,  corresponding to the  usual space-time components of the connection one-form,  and the scalar multiplet coming  from its matrix components $\phi = \phi_a \theta^a = \phi_a^b \lambda_b \theta^a$.   In the case when $\phi =0$, this lagrangian coincides with the one investigated in \cite{serie:03}, with $G=SU(n)$ and $R$ the defining representation. The complete analysis of general solutions of equations of motion seems to be too tedious at present. This is why we shall restrict ourselves to a qualitative analysis of the case when the space time components of $\Omega$ do vanish $F_{\mu \nu}=0$, leaving only the contribution of scalar multiplets degrees of freedom.  
\section{The reduced Lagrangian for scalar fields}

We now specialize in the case where the algebra is  $ C^{\infty}(\mathbb{R}^4) \otimes M_2(\mathbb{C})$, and choose the simplest ansatz with only one scalar field $ \varphi \in C^{\infty}( \mathbb{R}_4 ) $ :
\begin{align*}
 \phi &= \varphi \ \theta
\end{align*}

In this case, the determinant introduced in (\ref{binc})  is:
\begin{align*}
  \begin{vmatrix}
    \hat{g}_{\mu \nu} & i D\phi\\
    -iD\phi & \hat{g}_{ab}+i H 
  \end{vmatrix} \ ,
  \label{matrice}
\end{align*}
where 
 \begin{align*}
   H &=  \left\{\Omega_{ab} \right\}_{{a,b = 1,2,3}} &
   D\phi  &= \left\{ D_{\mu}\phi_a\right\}_{\substack{\mu=0,1,2,3 \\ a=1,2,3 \ \,  }} &
    \hat{g}_{\mu \nu}&=  g_{\mu \nu}\otimes  \mathbb{1}_2
   \ .
 \end{align*}

By virtue of Schur's lemma one can reduce this determinant to the one of the matrix:
\begin{equation*}
  \begin{vmatrix}
  \mathbb{1}_3   + i H - D_{\mu}{\phi}D^{\mu}\phi
    \end{vmatrix} \ .
\label{matrice:2}
\end{equation*}

It was shown in \cite{park:99,serie:03} that for matrices of this type, the determinant  is a perfect square. Therefore one can express its square root with a finite sum of traces of the matrix $M= i H - D\phi D\phi$ (from now on, we shall adopt the shortened notation  $  D_{\mu}{\phi}D^{\mu}\phi=D\phi D\phi =(D\phi)^2$), and finally one can deduce the most compact form of the desired lagrangian:
\begin{widetext}
\begin{align*}
  L&=1- \left\{ \left(1+ 3 \beta^{-2} (D\varphi)^2\right)^2 + 
    16 \beta^{-2} \varphi^2(\varphi-m)^2 \right\}^{\frac{1}{4}}
  \sqrt{1+4 \beta^{-2} \varphi^2(\varphi-m)^2 }  \ .
\end{align*}
\end{widetext}
where the parameter $\beta$, defining the critical field strength, and the parameter $m$, defining the mass  of the scalar field, hare explicitely displayed. In what follows they will be set to 1, in order to simplify the formulas.

This lagrangian is a particular case of the most general form of spherically symmetric ansatz, which contains one real and one complex scalar field in the noncommutative part and one complex scalar field and one $U(1)$ abelian gauge field  in the commutative part of the connection.

\section{The static field case}

In this section we show that there is no possibility to obtain non-trivial static configurations in the present system, with Higgs multiplet reduced to a single scalar. We generalize  Derrick's theorem \cite{derrick:64} to our particular lagrangian. The idea of the proof is to use  spatial dilatation of the field $\varphi (r) \to \varphi_{\lambda}(r)=\varphi (\lambda r) $ in order to generate a one-parameter curve in the space of fields around such a solution. Then the variational principle along this curve gives $\partial S[\varphi_{\lambda}] / \partial \lambda = 0$ at $\lambda=1$, i.e. : 
\begin{align}
\int 4 \pi  r^2 dr \left ( \frac{\partial \mathcal{L}}{\partial \varphi'} \varphi' - 3
 \mathcal{L} \right ) &= 0 \ .
\label{derrick}
\end{align} 
One can show that the function under the sum sign,
\begin{align*}
  f(\varphi, \varphi') &=1/3 \left ( \frac{\partial \mathcal{L}}{\partial
      \varphi'} \varphi' - 3 \mathcal{L} \right ) \notag \\
  &=  \frac{\sqrt{A}}{B^{3/4}} \left[(1+3p)(1+2p)+ 16 s^2 \right] -1 \ ,
  \label{eq:fonction} 
\end{align*} 
where
\begin{align*}
s &={\varphi} (\varphi-1) &
p &= \varphi'^2 \\
A &= 1+4 s^2 &
B&= (1+3p)^2+16 s^2  \ .
\end{align*}
Then $f$ is always non-negative. This can be easily seen if you observe the following implication:
\begin{align*}
  ((1+2p)(1+3p)+ 16 s^2)^4 \geq ( (1+3p)^2+ 16s^2)^3    \Rightarrow  f \geq 0 
\end{align*}
the first inequality being obvious. Therefore the condition (\ref{derrick}) is satisfied if and only if $f$  is zero for all values of $ r$. The  equation $f=0$ admits only trivial solutions,  $\varphi' = 0$ and $\varphi = 0 \text{ or } 1$. This  leads to the conclusion, as in the case originally considered by Derrick\cite{derrick:64}, that non-trivial solutions can not be found in this model.

\section{The time dependent scalar field}

A scalar field with an upper limit on its strength can be used as driving force in primordial cosmological models. This is why an homogeneous, only time dependant configuration should be studied before being coupled to gravitation.

We have performed numerical analysis of the time dependent configurations of scalar field resulting from the simplest ansatz $\varphi = \varphi(t)$. It gives an interesting phase space portrait and  confirms the idea that  Born-Infeld-like  theories set an upper bound on the velocities ( i.e. time derivatives of  $\varphi$), and on the field strength as well. Such an ansatz could be of  interest in cosmological models, when it can be coupled with the scale factor $a(t)$ of Robertson-Friedmann metric. The equations of motion in this case take on the following form :
\begin{align*}
  &\dot{\varphi}=u\\
  &(1+4X) g(X,Y) \dot{u} + 4 s s' h(X,Y) = 0 \ ,
\end{align*}
where
\begin{align*}
  & s=\varphi (\varphi -1) \ , \ s'=2\varphi-1\\
  &  X=s^2 \ ,  \ Y=u^2\\
  & g(X,Y)= 16 X (1-9 Y ) + (1-3 Y)^2 \\
  & h(X,Y)= ((1-3Y)^2+16 X)(1-Y+8X) -6(1+4 X)(1-3Y)Y  
\end{align*}

\begin{figure}[!h]
\centering
\includegraphics[]{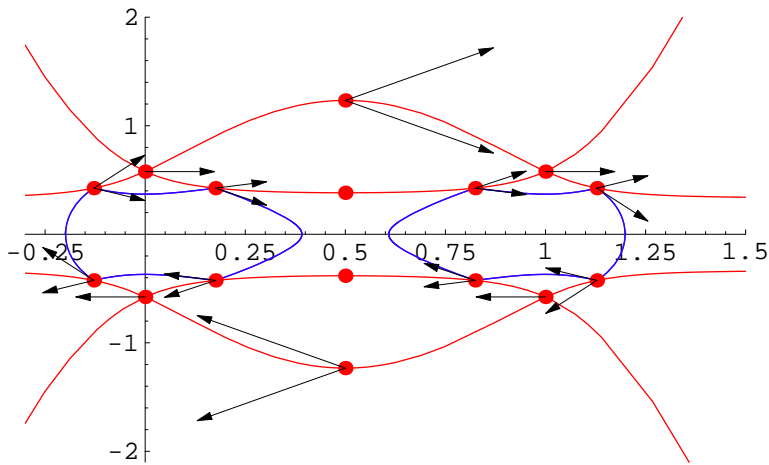} 
\caption{Characteristic curves and points in the phase space}
\label{char}
\end{figure}

At some points of the phase space the derivative $\dot{u}$ is not well defined. These are the points at which the polynomial $g$ vanishes (4 curves in Fig~\ref{char}). Nevertheless, at certain points (where $ s s' h(X,Y)=0$) the singular behavior is only apparent, because the undetermined ratios $0/0$ happen to have finite limits. The total number of these particular points in the phase space is $16$, but only $2$  of them display genuine singularity.  In the $14$ remaining special points, $\dot{u}$ is kept finite. For these points it is possible to calculate  analytically the tangent vector of limiting trajectories passing through them (see vectors in Fig.~\ref{char}). The only two points with genuine singularity are the ones without any vector attached to them. They are found on the central vertical line $\varphi = 0.5$ on both sides of the horizontal line and close to it.

\begin{figure}[!h]
\centering
 \includegraphics[]{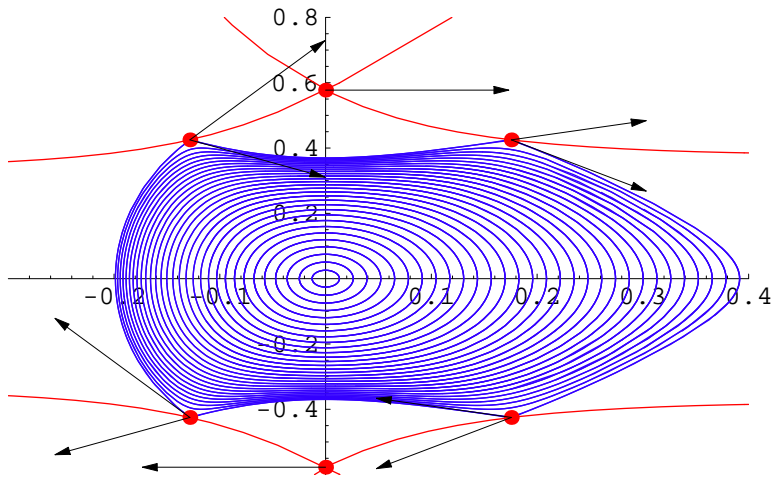}
\caption{Trajectories  in  the confined region of the phase space}
\label{traj}
\end{figure}

The phase space portrait is symmetric with respect to reflections around the vertical line $\varphi = 0.5$. Cyclic trajectories are contained inside two pentagon-like areas circumscribed by separatrices. These areas are disposed symmetrically with respect to the vertical line $\varphi = 0.5 $. One of these areas is represented in more details  in Fig.~\ref{traj}.
One can note that in a certain region of the phase space the trajectories are periodic and are defined for all values of time $t$. If one chooses the initial conditions outside this region, integration ends up after some finite time. This means that the solutions $\varphi(t)$ obtained with these initial conditions have their second derivative divergent after finite time when they hit one of the curves on which $g = 0$.

Nevertheless some of these curves, with very special initial conditions, can go beyond the singular curve $g = 0$ passing through the points at which the  undefinite expressions become finite again. 

\section{Conclusion and perspectives}

In this article we have proposed another generalization of the so-called Dirac-Born-Infeld lagrangian (see \cite{tseytlin:97,gibbons:98}). Next,  we have studied some  solutions in the case of simplest ansatz with 2 by 2 matrices. We have  first considered  solutions in 3-dimensional space with spherical ansatz and proved a no-go theorem (similar to the one found by  Derrick \cite{derrick:64}).  We then concentrated our attention on the 1-dimensional case (pure time-dependent) and investigated the dynamical properties of the present model.

The highly non-linear behavior of the field $\varphi$ in this model suggests that when coupled to gravitation in a standard way, i.e. via minimal coupling resulting from the replacement of ordinary derivatives by their covariant counterparts in presence of the Einstein-Hilbert lagrangian for gravitational field, it may lead to unusual behavior of cosmological models. The investigation of such models using this scalar field will be the subject of a forthcoming paper \cite{troisi:04}.

\bibliography{biblio_articles}
\bibliographystyle{h-physrev4}

\end{document}